\documentclass[sigconf,9pt,natbib=false]{acmart}

\renewcommand\footnotetextcopyrightpermission[1]{} 
\RequirePackage[abbreviate=true, dateabbrev=true, isbn=true, doi=true, urldate=comp, url=true, maxbibnames=9, backref=false, backend=biber, style=ACM-Reference-Format, language=american]{biblatex}

\usepackage{graphicx}

\PassOptionsToPackage{bookmarks=false}{hyperref}
\usepackage{booktabs} 
\usepackage{multirow} 

\newcommand{\ToolName}{\textsc{\textsc{ProbLP}}}
\newcommand{\Fx}[1]{\tilde{#1}}
\newcommand{\FxW}[1]{\overset{\sim}{#1}_{\text{fx}}}
\newcommand{\FlW}[1]{\overset{\sim}{#1}_{\text{fl}}}

\newcommand{\CustSpace}[0]{10pt}
\newcommand{\EqSize}{\normalsize}

\usepackage{amsthm}
\theoremstyle{definition}

\usepackage{subcaption}

\setcopyright{None}

\usepackage{pifont} 

\begin{document}

\title{\ToolName{}: A framework for low-precision probabilistic inference}

\author{Nimish Shah}
\orcid{0000-0003-3234-0715}
\affiliation{%
  \institution{MICAS, Electrical Engineering, KU Leuven}
}
\email{nshah@esat.kuleuven.be}

\author{Laura I. Galindez Olascoaga}
\affiliation{%
  \institution{MICAS, Electrical Engineering, KU Leuven}
}
\email{laura.galindez@kuleuven.be}

\author{Wannes Meert}
\affiliation{%
  \institution{DTAI, Computer Science, KU Leuven}
}
\email{wannes.meert@kuleuven.be}

\author{Marian Verhelst}
\affiliation{%
  \institution{MICAS, Electrical Engineering, KU Leuven}
}
\email{marian.verhelst@kuleuven.be}

\begin{abstract}
Bayesian reasoning is a powerful mechanism for probabilistic inference in smart edge-devices. During such inferences, a low-precision arithmetic representation can enable improved energy efficiency. However, its impact on inference accuracy is not yet understood. Furthermore, general-purpose hardware does not natively support low-precision representation. To address this, we propose \ToolName{}, a framework that automates the analysis and design of low-precision probabilistic inference hardware. It automatically chooses an appropriate energy-efficient representation based on worst-case error-bounds and hardware energy-models. It generates custom hardware for the resulting inference network exploiting parallelism, pipelining and low-precision operation. The framework is validated on several embedded-sensing benchmarks.

\end{abstract}


\maketitle

\section{Introduction}
The use of probabilistic inference is popular for robust classification, diagnosis and decision-making problems, because of its ability to assign a confidence-level to every result, in terms of probability. Probabilistic Graphical Model (PGM) \cite{Pearl1988PRIS}, an established tool for probabilistic inference, is widely used for such problems. PGMs have several interesting properties that make them suitable for embedded applications. Specifically, PGMs: 1) are capable of dealing with missing data; 2) allow to incorporate information from different domains, as well as expert knowledge; 3) can be trained with less data; and 4) can explicitly model uncertainty and causal relationships in the system. In addition, PGMs' performance is competitive with respect to other state-of-the-art Machine Learning implementations on embedded sensing applications \cite{galindez2018dynamic,jonas,george2017generative,Liang2019AAAI}.

Inference in PGMs is prominently performed using a versatile representation known as an Arithmetic Circuit (AC) (or sum-product network) \cite{CHAVIRA2008772}. An AC refers to a model of computation and is often represented as a graph of additions and multiplications. ACs allow for an integration of both statistical and symbolic methods in artificial intelligence, a promising combination that is pursued in state-of-the-art machine learning methods \cite{Thompson2018Wired,Manhaeve2018NIPS,Liang2019AAAI}. They are also central to performing inference in the field of probabilistic (logic) programming \cite{Fierens2015TPLP,Manhaeve2018NIPS}. Furthermore, recent approaches learn ACs directly from data, with state-of-the-art performance in certain applications \cite{Liang2019AAAI}. In this work, we focus on ACs representing Bayesian networks (BN), a type of PGM.

 Inference in ACs is generally restricted to obtaining exact solutions on general purpose computing devices. A significant improvement in energy efficiency would be possible by tolerating some error through approximating the probability computed by ACs. Take for instance a smartphone-based activity identification for elderly, wherein a probability is evaluated for different activities (e.g., a user walking up the stairs). The application chooses to identify an activity only if its probability is higher than a certain threshold, say 0.60. Here, allowing an output error of 0.01 would only affect the decisions within the probability range of 0.59 and 0.61, while enabling improved energy-efficiency.

A promising hardware optimization that can exploit the available error-tolerance is to realize the additions and multiplications in reduced-precision representation. Yet, the state-of-the-art is lacking analysis of the impact of such precision reduction on the output probability error. Previous works \cite{chan2002numbers,chan2004sensitivity,tschiatschek2015bayesian} have studied the impact of low-precision in leaf nodes of an AC, but do not account for noisy or low precision computations in its internal nodes. However, the error in the imprecise internal nodes can get accumulated and be the dominant source of imprecision in the inference output. 

In this paper, we propose \ToolName{}\footnote{Code available at https://github.com/nimish15shah/ProbLP}, a holistic framework to automate the design of low-precision energy-efficient hardware for probabilistic inference in Arithmetic Circuits. Our contributions are as follows: 
\begin{itemize}
    \item We derive bounds on the error in probabilistic queries due to low-precision representation, taking into account the error introduced in all the nodes in an AC.
    \item We develop energy models to help choose the most energy-efficient representation
    \item We develop a tool to automatically generate low-precision inference hardware, and validates its performance on several embedded sensing benchmarks. 
\end{itemize}

This paper is organized as follows. Section \ref{sec:background} gives an introduction to Arithmetic circuits compiled from Bayesian Networks and an overview of related works. In Section \ref{sec:method}, we derive analytical error bounds for ACs. We introduce the \ToolName{} framework and elaborate on how it selects the optimal precision and selects between fixed- or floating-point representation. Section \ref{sec:experiments} demonstrates the validity of the framework on a suite of embedded sensing benchmarks and Section \ref{sec:conclusion} concludes this work.

\setlength{\abovecaptionskip}{10pt}
\begin{figure}[!t]
\centering

\begin{subfigure}[b]{0.24\textwidth}
    \centering
    \includegraphics[width=\columnwidth]{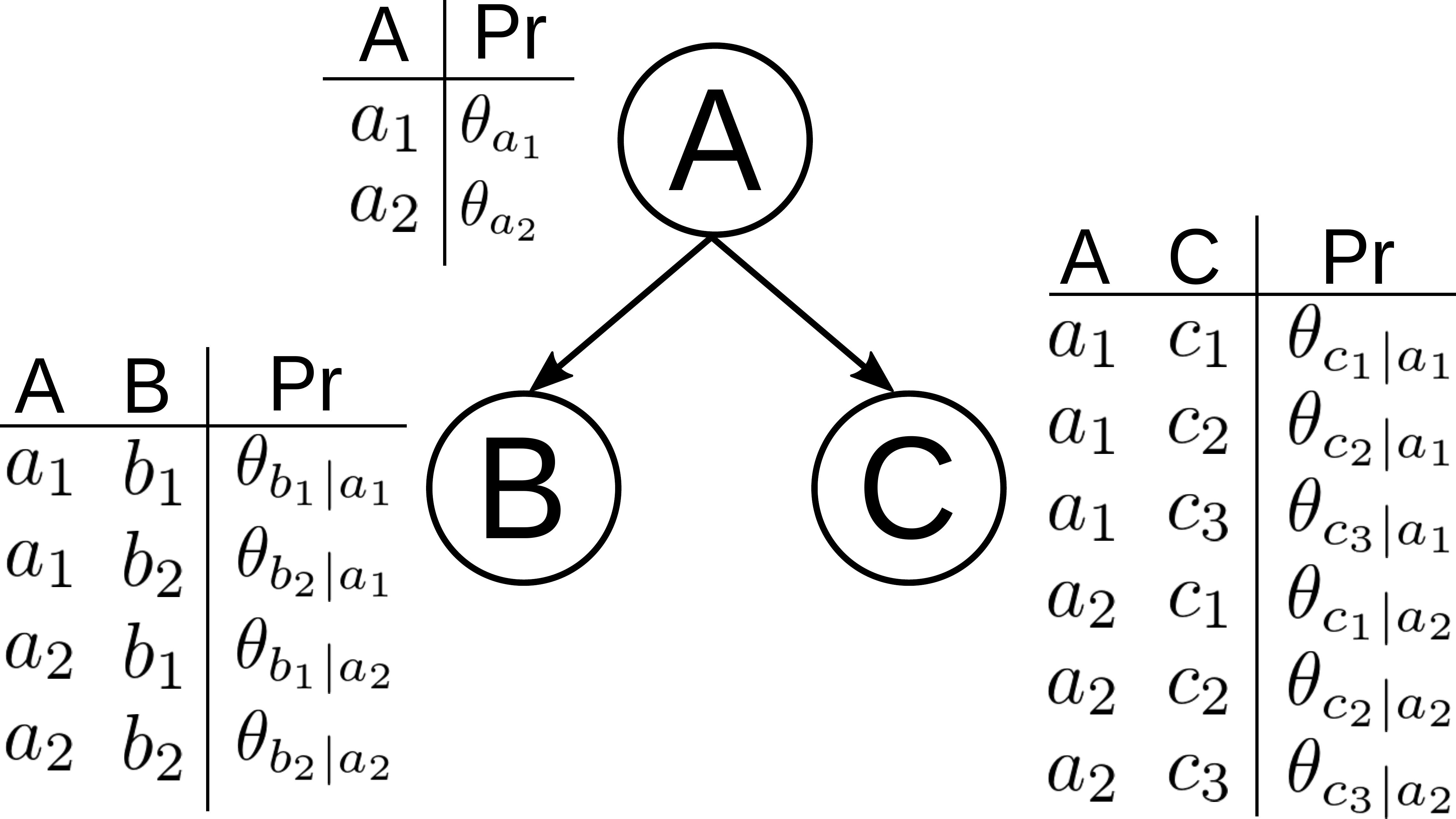}
    \caption{ }
    \label{ExampleBN}
\end{subfigure}
\vspace{0.3cm}

\begin{subfigure}[b]{0.48\textwidth}
\centering
\includegraphics[width=0.9\columnwidth]{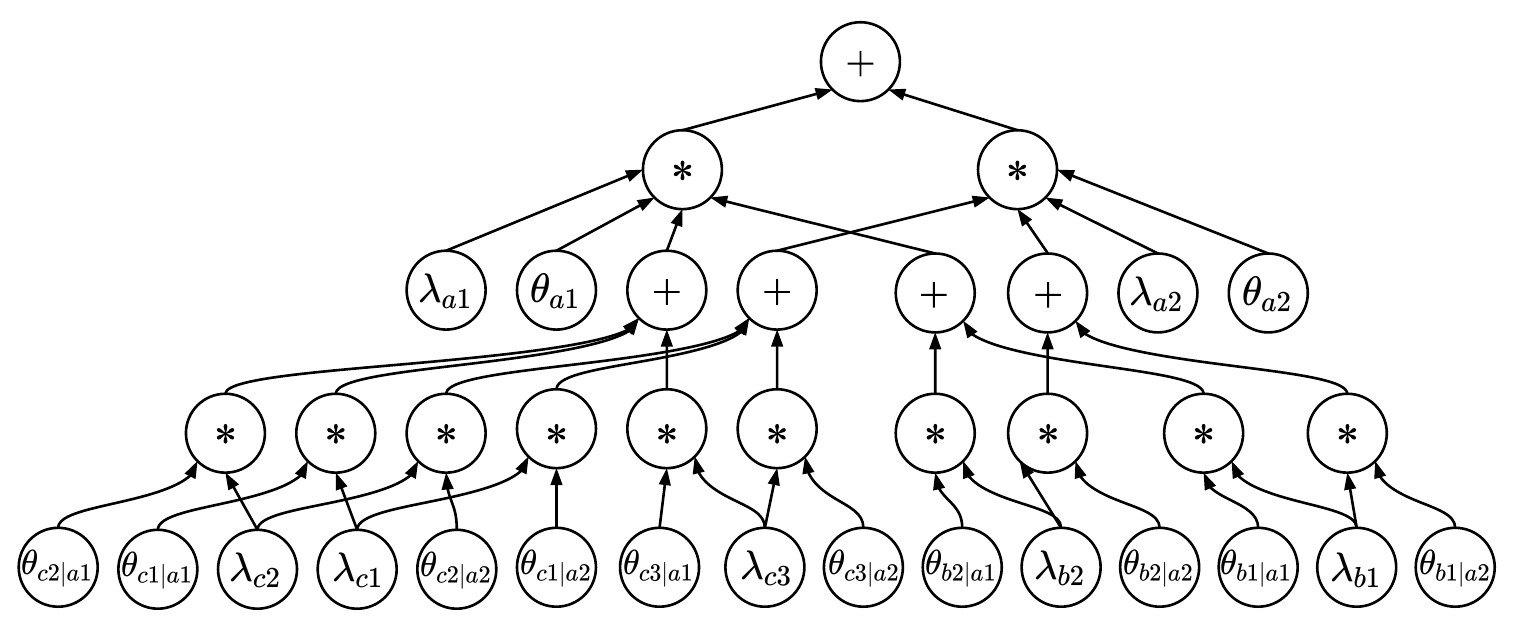}
\caption{ }
\label{ExampleAC}
\end{subfigure}
\caption{(a) Example of a Bayesian Network and (b) Arithmetic compiled from the BN of Figure 1a.}
\end{figure}
\setlength{\abovecaptionskip}{5pt}

\section{Background and previous work} \label{sec:background}
In this paper, we denote a random variable with uppercase letter $X$ and its instantiation with a lower case letter $x$. A set of multiple random variables are denoted with bold upper case letters $\textbf{X}$ and its joint assignment with bold lower case letters $\textbf{x}$. 

Bayesian networks (BN) are directed acyclic graphs that compactly encode a joint probability distribution over a set of random variables $\{X_1,...,X_n\}$ \cite{Pearl1988PRIS}:%
{\EqSize
\begin{align} \label{joint}
Pr(X_1,...,X_n)=\prod_{\substack{i=1}}^{n} Pr(X_{i}|\boldsymbol{\Pi}_{X_{i}}),
\end{align}
}%

where $\boldsymbol{\Pi}_{X_i}$ denotes the parents of $X_i$ and $Pr(X_{i}|\boldsymbol{\Pi}_{X_{i}})$ are the conditional dependencies between variables and their parents, which can be represented as Conditional Probability Tables (CPTs). In the graphical component of BNs, the variables are represented as nodes and their probabilistic or causal relationships are indicated by the direction of the edges among them, as depicted in Figure \ref{ExampleBN}. The joint probability distribution in (\ref{joint}) allows answering a number of probabilistic queries such as the marginal probability, the conditional probability or the Most Probable Explanation (MPE)\cite{Pearl1988PRIS}. 

Probabilistic inference on a BN can be made efficient by compiling it to an Arithmetic circuit, which consists only of multiplications and addition. Figure \ref{ExampleAC} shows an example of an AC generated by compiling the BN in Figure \ref{ExampleBN}. 
The inputs to this AC are the BN's parameters, represented by $\theta_{x|u}$, where $x$ and $u$ are the instantiation of random variable $X$ and its parents. The second type of inputs to the AC are indicators $\lambda_{x}$, which are binary variables that indicate the evidence of the observed nodes. 
The probability of an evidence (e.g., $\mathbf{e}\,$=$\,\{A\,$=$\,a_1,C\,$=$\,c_3\}$) can be computed with an upward pass on the AC by setting the indicators that contradict the evidence to 0 ($\lambda_{a_2}$= $\lambda_{c_1}$= $\lambda_{c_2}$=$\,0$), and others to 1 ($\lambda_{a_1}$= $\lambda_{b_1}$= $\lambda_{b_2}$= $\lambda_{c_3}$=$\,1$).  

Previous works have studied the impact of finite-precision CPT parameters on marginal and conditional probability \cite{chan2002numbers,chan2004sensitivity,tschiatschek2015bayesian}. This helps to reduce memory footprint due to a smaller inference model. However, they did not study the effect of low-precision arithmetic operations. The authors of 
\cite{zermani2015fpga,khan2016hardware} studied the effect of fixed-point arithmetic on marginal probability for a few BNs, but not of conditional probability, and did not provide error bounds. Moreover, the impact under floating-point arithmetic operation is also unclear.

This work provides analytical bounds on the absolute and the relative error in marginal and conditional probabilities for fixed- and floating-pt operations in the entire AC. A holistic framework \ToolName{} is introduced, which also takes energy-consumption into account to choose the optimal representation among fixed-point and floating-point. Subsequently, it automatically generates custom hardware for the AC evaluation. 

\section{Methodology} \label{sec:method}
\begin{figure}[!t]
\centering
\includegraphics[width=\columnwidth]{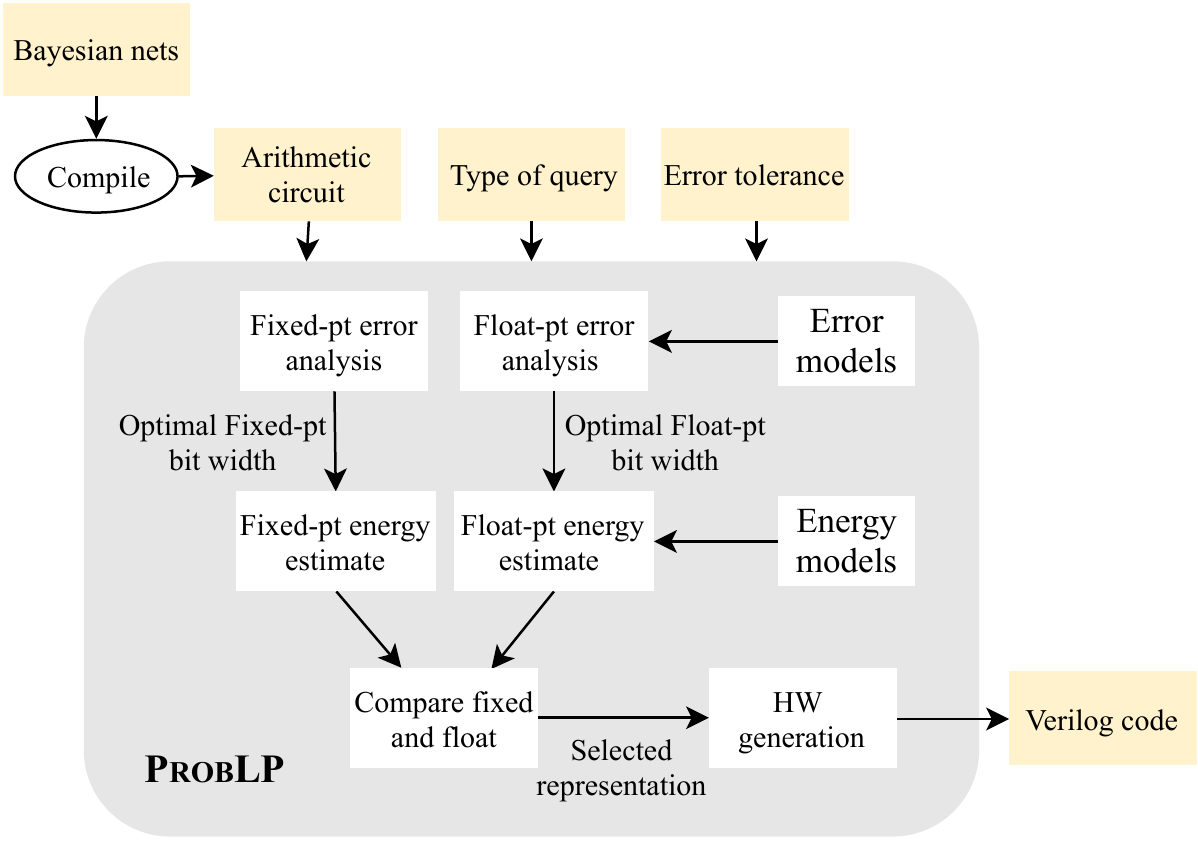}
\caption{Internal working of \ToolName{}}
\label{fig:ToolFlow}
\end{figure}

Different components of the \ToolName{} framework are shown in figure \ref{fig:ToolFlow}. \ToolName{} takes in an AC together with some user requirements, based on which, it calculates the least number of fixed and floating point bits needed to meet these requirements. To do so, \ToolName{} evaluates error-bounds for the AC, based on their error models.
To choose between these two representations, it subsequently estimates the energy of the complete AC based on energy models. Finally, it generates a fully-parallel pipelined hardware in the selected low-precision representation. \\
The three inputs of \ToolName{} are as follows:

\textbf{Arithmetic circuit}: The Arithmetic circuit to be implemented using low-precision hardware. In this paper, we use ACs compiled from Bayesian networks, but they can as well be compiled from probabilistic (logic) programs or can be trained directly from data.

\textbf{Type of query}: The type of probabilistic query to be performed using the AC, to be chosen from \emph{marginal probability}, \emph{conditional probability} or the probability of \emph{most probable explanation} (MPE). 

\textbf{Error tolerance}: The amount of error on the output that can be tolerated in the probabilistic queries by the application, for \emph{all} possible combination of inputs, in terms of \emph{absolute} or \emph{relative} error. An absolute error is given as $\epsilon= \tilde{Pr} - Pr$ and a relative error is given as $\epsilon= \frac{\tilde{Pr} - Pr}{Pr}$, where $Pr$ is the output probability of interest.%

\subsection{Error analysis} \label{sec:error_estimation}
The aim of error analysis is to estimate the minimum number of bits required to achieve the user-specified error tolerance. For this, it has to take into account the impact of reducing the number of bits on the error in the AC output probability. There are two sources of error in an AC: an error in the leaf nodes when CPT values are quantized to finite precision, and an error injected in the intermediate nodes of AC due to the finite precision arithmetic operations. Unlike previous research works,
we formally treat the error in the intermediate nodes as well to derive the error-bounds.

We consider two representations: fixed-point and floating-point. Operators of both types round bits during computation. For example, a multiplication of 2 inputs of $n$ bits produces an exact result with $2n$ bits, which is subsequently rounded to fit an $n$ bit output.The error introduced can be modeled as an additive noise source. 

The error models used for the leaf nodes and the intermediate nodes are described next. Some of the models are inspired by \cite{higham2002accuracy}, but authors of that work did not perform the error analysis for ACs, and some of the models would render unbounded errors if AC-specific constraints are not exploited. The arithmetic operators are assumed to round the extra bits to the nearest value. 
\subsubsection{Fixed-pt error estimation} \label{FxPt_models}
Let $I$ and $F$ be the number of integer and fraction bits. All the numbers are assumed to be in the range of the fixed-pt format, implying an absence of overflow during computation, which can be ensured by using an appropriate number of integer bits $I$, discussed in detail in section \ref{IntExpBits}.

\textbf{Fixed-pt leaf node}:
Let $a$ be the real value of a leaf node in an AC, and $\Fx{a}$ be its fixed-pt representation. The error in fixed-pt conversion can be bounded as,
{\EqSize
\begin{align} 
\left| \Delta a \right| = \left|\Fx{a} - a\right| \le 2^{-(F+1)} \label{Fxleaf}
\end{align}
}%

\textbf{Fixed-pt adder node}:
If $\Fx{a}$ and $\Fx{b}$ are the fixed-pt representation of adder inputs $a$ and $b$, the error in the output $\Fx{f}$ is given as
{\EqSize
\begin{align} \label{FxAdd}
\Delta f &= \Fx{f} - f = (\Fx{a} + \Fx{b}) - (a+b) = (a + \Delta a + b + \Delta b) - (a+b) \nonumber \\ 
& = \Delta a + \Delta b
\end{align}
}%

Note that the fixed-pt adder does not add any error of its own, as it does not round bits, and hence simply accumulates the error of the inputs. Note again that the adder output cannot overflow, as all the numbers are ensured to be in range.

\textbf{Fixed-pt multiplier node}: 
With $\Fx{a}$ and $\Fx{b}$ as the fixed-pt representation of multiplier inputs $a$ and $b$, the error in fixed-pt multiplier output $\Fx{f}$ can be bounded as,%
{\EqSize
\begin{align}
\Delta f & = \Fx{f} - f \le \left(\widetilde{ab} + 2^{-(F+1)}\right)- ab   \label{eq:FxMul_err} \\ 
& = (a + \Delta a)(b+ \Delta b) - ab + 2^{-(F+1)} \nonumber \\
&\le a_{\textrm{max}}\Delta b + b_\textrm{{max}} \Delta a + \Delta a \Delta b +2^{-(F+1)} \label{FxMul}
\end{align}
}%

In (\ref{eq:FxMul_err}), the error term $2^{-(F+1)}$ models the error introduced when the LSB bits of the intermediate multiplication result are rounded to fit back into $F$ fractional bits. Equation (\ref{FxMul}) produces an unbounded error unless $a_\textrm{{max}}$ and $b_\textrm{{max}}$ can be bounded. 

The $a_\textrm{{max}}$ and $b_\textrm{{max}}$ can be efficiently bounded by taking into account the AC-specific properties. An AC consists of adders and multipliers and only operates on non-negative numbers. As a result, each internal node in the AC is a monotonously increasing function of its inputs. Hence, all the nodes are at the maximum value when all the inputs are at their maximum. As such, since CPT parameters stay constant across AC evaluations, 
this is achieved when all the indicator variables $\lambda_{x}$ are set to 1. This allows to assess the $a_\textrm{{max}}$ and $b_\textrm{{max}}$ of every operator in the AC with just a single AC evaluation. Thereby, allowing \ToolName{} to bound the error of fixed-pt multipliers.


\vspace{\CustSpace}
\subsubsection{Floating-pt error estimation} \label{FltPt_models}
Let $E$ and $M$ be the exponent and mantissa bits. We only consider normalized floating-pt here. All the numbers are assumed to be within the range of the given format, ensured by a method explained in detail in section \ref{IntExpBits}.

\textbf{Float-pt leaf node}:
Let $a$ be the real value of a leaf in AC and $\Fx{a}$ be its floating-pt representation. The absolute error introduced due to the floating-pt conversion can be bounded as described in \cite{higham2002accuracy},%
{\EqSize
\begin{align} \label{FlLeaf}
\left| \frac{\Delta a}{a} \right| \le 2^{-(M+1)}
\end{align}
}%
which can be expressed alternatively as,%
{\EqSize
\begin{align} \label{FlLeaf_alternate}
\Fx{a} & = a + \Delta a = a\left(1 + \frac{\Delta a}{a}\right) = a \left(1 \pm \epsilon \right) 
\end{align}
where: \vspace{-0.2cm}
\begin{description}
\item{$0 \le \epsilon \le 2^{-(M+1)}$}
\end{description}
}%

\textbf{Float-pt adder node}:
Let $\Fx{a}$ and $\Fx{b}$ be the float-pt versions of adder inputs $a$ and $b$, $f$ be the ideal output, and $\Fx{f}$ be the output of a floating-pt adder. $\Fx{a}$ and $\Fx{b}$ can be represented as,%
{\EqSize
\begin{align} 
\Fx{a} &= a \left(1 \pm \epsilon \right) ^ m \;\;\textrm{  and  }\;\; \Fx{b} = b \left(1 \pm \epsilon \right) ^ n \label{Flinputs}
\end{align}
}%
Here, $m$ and $n$ depends on the amount of error accumulated in $a$ and $b$, respectively. The bound on $\Fx{f}$ can be given as follows,%
{\EqSize
\begin{align} 
\Fx{f} &= \left(\Fx{a} + \Fx{b}\right) \left(1 \pm \epsilon \right) = \left( a \left(1 \pm \epsilon \right) ^ m + b \left(1 \pm \epsilon \right) ^ n \right) \left(1 \pm \epsilon \right) \label{FlAdder_inherent} \\
&= \left(a + b \right)\left(\frac{a}{a+b} \left(1 \pm \epsilon \right) ^ m + \frac{b}{a+b} \left(1 \pm \epsilon \right) ^ n \right) \left(1 \pm \epsilon \right) \nonumber \\
&= f\left(\frac{a}{a+b} \left(1 \pm \epsilon \right) ^ {m+1} + \frac{b}{a+b} \left(1 \pm \epsilon \right) ^ {n+1} \right) \nonumber \\
& \le 
\begin{cases}
f(1 \pm \epsilon) ^ {m+1} & m>n \\
f(1 \pm \epsilon) ^ {n+1} & \textrm{otherwise}
\end{cases}
\label{FlAdder}
\end{align}
}%

The error term in (\ref{FlAdder_inherent}) is due to the rounding of LSB bits of the mantissa of the smaller input before addition.

\textbf{Float-pt multiplier node}:
Just as in case of the adder, the inputs $\Fx{a}$ and $\Fx{b}$ can be bounded as in (\ref{Flinputs}). With that, the output of a floating-pt multiplier can be given as,%
{\EqSize
\begin{align} 
\Fx{f} &= \widetilde{ab}\left(1 \pm \epsilon \right) = ab \left(1 \pm \epsilon \right) ^ m \left(1 \pm \epsilon \right) ^ n \left(1 \pm \epsilon \right) \label{FlMul_inherent} \\
&= f\left(1 \pm \epsilon \right) ^ {m+n+1} \label{FlMul}
\end{align}
}%

The error term in (\ref{FlMul_inherent}) is due to the rounding of the LSB bits of the mantissa to fit the result in $M$ mantissa bits.
\vspace{\CustSpace}
\subsubsection{Error-bound at the AC output} \label{sec:ErrBound_ACout}
\begin{figure}[!t]
\centering
\includegraphics[trim={0cm 1.5cm 0cm 1.5cm},clip, width=0.8\columnwidth]{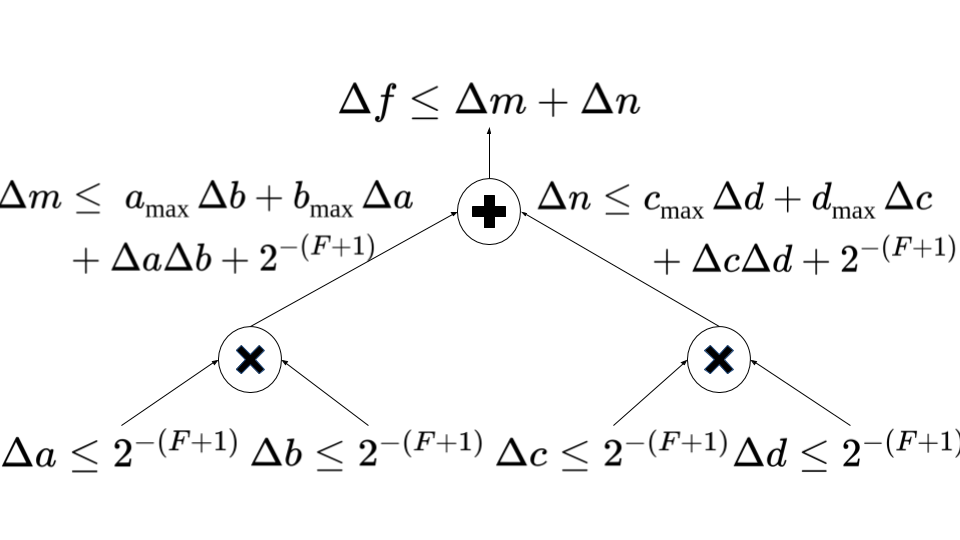}
\caption{Error propagation with fixed-pt error models.}
\label{ErrProp}
\end{figure}

Equations (\ref{Fxleaf}), (\ref{FxAdd}), (\ref{FxMul}), and (\ref{FlLeaf}), (\ref{FlAdder}) and (\ref{FlMul}) corresponds to the \textit{Error models} shown in figure \ref{fig:ToolFlow}. As these models generate the output in the same format as the inputs, they provide a way to recursively propagate the error from the leaves of an AC all the way up to its output node, by accumulating the error introduced in every adder and multiplier. Figure \ref{ErrProp} shows an example of error propagation using the fixed-pt error models. This is performed as a part of the \textit{fixed-pt error analysis} and \textit{float-pt error analysis} blocks of \ToolName{} shown in figure \ref{fig:ToolFlow}.

The error propagation in fixed-pt arithmetic produces a bound of the form $\Delta f \le c$, where $\Delta f$ is the absolute error in the output node, and $c$ is a constant that depends on the size and structure of the AC, its parameters, and the number of fixed-pt bits. The constant $c$ can be estimated recursively with our error models for any given AC. Similarly, the error propagation in floating-pt arithmetic produces a bound of the form $\Fx{f} \le f\left(1 \pm \epsilon \right) ^ {c}$, where $\Fx{f}$ is the output of an AC with floating-pt operators, $f$ is the ideal output, $\epsilon$ is a constant related to the number of floating-pt bits, and $c$ is a constant related to the size and structure of the AC. Again, the constant $c$ can be estimated recursively using the models we proposed, for any AC. Alternatively, the floating-pt bound can be expressed as $\frac{\tilde{f} - f}{f} \le \delta$ for some constant $\delta$, i.e., a bounded relative error at the output. 

\subsubsection{Number of integer or exponent bits} \label{IntExpBits}
For the error-models proposed in section \ref{FxPt_models} and \ref{FltPt_models} to be valid, the numbers encountered during the computation must stay within range of the representation. This can be ensured by using an appropriate number of integer bits $I$ and exponent bits $E$ for fixed- and floating-pt respectively. Otherwise, error in some of the probability evaluations would exceed the predicted bounds. It is hence important to automatically derive the required range of numbers for any given AC.

\textbf{Max-value analysis}: The largest number to be encountered in an AC can be derived by setting all the indicator variables $\lambda_{x}$ to 1, as explained in section \ref{FxPt_models}. Analyzing the internal AC data values of this query, 
allows deriving the required $I$, resp. $E$ to avoid overflow.

\textbf{Min-value analysis}: The floating-pt models are invalid in case of underflow as well. Hence, it is necessary to also estimate the smallest positive non-zero value for an AC. It can be proven that all the nodes in an AC are at the respective minimum non-zero values when all the indicator variables $\lambda_{x}$ are set to 1 and the adders are replaced with minimum operators $min\left(i,j\right)$. 
The resulting efficient AC evaluation allows \ToolName{} to analyze a lower bound on AC values, and find the appropriate $E$ required to prevent underflow. The fixed-pt models remain valid even if the number of fraction bits is not enough to represent small values in the AC, so no special precautions are needed here regarding underflow.

In this way, \ToolName{} performs the Max-value and Min-value analysis to selects $E$, resp. $I$, that satisfies both the requirements. 

\subsection{Bounds for probabilistic queries} \label{sec:bounds_prob_queries}
As shown in figure \ref{fig:ToolFlow}, \ToolName{} aims to estimate the optimal fixed-pt and float-pt bit width for a given type of probabilistic query and error tolerance. However, the bounds derived so far, apply only to a single AC evaluation. Some types of probabilistic queries require a combination of multiple AC evaluations. In this section, we derive bounds for two type of probabilisitic queries: 1) Marginal probability and MPE, and 2) Conditional probability.
\subsubsection{Marginal probability and MPE} \label{para:MargProb}
Marginal probabilities $Pr(\textbf{q},\textbf{e})$ and most probable explanation (MPE) need only one AC evaluation. Hence, the bounds derived in section \ref{sec:ErrBound_ACout} apply for these queries.
\subsubsection{Conditional probability} \label{sec:CondProb_bound}
Conditional probability $Pr(\textbf{q}\,|\,\textbf{e})$ is evaluated by performing two AC evaluations, 
one for $Pr(\textbf{q},\textbf{e})$ and one for $Pr(\textbf{e})$, followed by taking the ratio of the two results\footnote{$Pr(\textbf{q}\,|\,\textbf{e})$ can also be estimated by an upward and a downward pass in an AC followed with a division. We do not consider it explicitly, but similar error bounds are expected.}. 

\textbf{Fixed-pt bounds}: In the case of fixed-pt arithmetic, the absolute error in each of the AC queries remains bounded. The impact on the conditional probability can hence be given as,%
{\EqSize
\begin{align}
\FxW{Pr}{\left(\textbf{q}\,|\,\textbf{e}\right)} &= \frac{\FxW{Pr}\left( \textbf{q}, \textbf{e} \right) }{\FxW{Pr}\left(\textbf{e}\right)} = \frac{ Pr\left( \textbf{q}, \textbf{e} \right) + \Delta_1}{Pr\left(\textbf{e}\right) + \Delta_2} 
\end{align}
}%

Here, maximum error is achieved when $\Delta_2=0$ and $\Delta_1=\Delta_{1\text{max}}$. In such a case, following equations show the impact on absolute and relative error in the conditional probability,%
{\EqSize
\begin{align} 
& \FxW{Pr}\left(\textbf{q}\,|\,\textbf{e}\right) = Pr\left(\textbf{q}\,|\,\textbf{e}\right) + \frac{\Delta_{1\text{max}}}{Pr\left(\textbf{e}\right)} \nonumber \\ 
 \implies & \Delta Pr_{\text{fx}}\left(\textbf{q}\,|\,\textbf{e}\right) = \frac{\Delta_{1\text{max}}}{Pr\left(\textbf{e}\right)} \le \frac{\Delta_{1\text{max}}}{min \; Pr\left(\textbf{e}\right)}  \label{CondQueryAbs} \\
 \implies & \frac{\Delta Pr_{\text{fx}}\left(\textbf{q}\,|\,\textbf{e}\right)}{Pr\left(\textbf{q}\,|\,\textbf{e}\right)} = \frac{\Delta_{1\text{max}}}{Pr\left(\textbf{e}\right)Pr\left(\textbf{q}\,|\,\textbf{e}\right)} \label{eq:CondQueryRel_Fx}
\end{align}
}%

Equations (\ref{CondQueryAbs}) and (\ref{eq:CondQueryRel_Fx}) show the absolute and relative error in the conditional query $Pr\left(\textbf{q}\,|\,\textbf{e}\right)$.
The error in the numerator is scaled by $Pr\left(\textbf{e}\right)$ and $Pr\left(\textbf{e}\right)$ $Pr\left(\textbf{q}\,|\,\textbf{e}\right)$, and these probabilities can become very small. Hence, large number of fixed-pt bits are generally required to achieve a reasonable error-bound, especially for the relative-error bound of (\ref{eq:CondQueryRel_Fx}). The absolute-error bound in (\ref{CondQueryAbs}) can be quantified by estimating the minimum possible value for $Pr\left(\textbf{e}\right)$ as described in section \ref{IntExpBits}, wherein adders are replaced with min operators. 

As the denominator of (\ref{eq:CondQueryRel_Fx}) can become very small, it is not a good idea to use fixed-pt when requiring a relative error bound in conditional probabilities. Moreover, quantifying a bound for (\ref{eq:CondQueryRel_Fx}) is also not straightforward. Hence, \ToolName{} will always choose float-pt for relative error in conditional probability. 

\textbf{Float-pt bounds}: The impact of using float-pt arithmetic on conditional probability can be given as follows.%
{\EqSize
\begin{align} \label{CondQueryRel}
\FlW{Pr}{\left(\textbf{q}\,|\,\textbf{e}\right)} &= \frac{\FlW{Pr}\left( \textbf{q}, \textbf{e} \right) }{\FlW{Pr}\left(\textbf{e}\right)} = \frac{ Pr\left( \textbf{q}, \textbf{e} \right)\left( 1\pm\epsilon \right)^m}{Pr\left(\textbf{e}\right)\left( 1\pm\epsilon \right)^n} \nonumber \\
&= Pr\left(\textbf{q}\,|\,\textbf{e}\right) \left( 1\pm\epsilon \right)^{m-n}
\end{align}
}%

In (\ref{CondQueryRel}), $m$ and $n$ are upper bounded by a constant, say $c$, but not lower bounded. In the worst case, one of them can become 0, while the other is $c$. Even in this worst case, the floating-pt version of the conditional probability still remains bounded as follows.%
{\EqSize
\begin{align}
Pr\left(\textbf{q}\,|\,\textbf{e}\right) \left( 1\pm \epsilon \right)^{-c} 
\le \FlW{Pr}\left(\textbf{q}\,|\,\textbf{e}\right)
\le Pr\left(\textbf{q}\,|\,\textbf{e}\right) \left( 1\pm\epsilon \right)^{c}
\end{align}
}%

This ensures a bound on the relative error $\frac{\Delta Pr_{\text{fl}}\left(\textbf{q}\, |\, \textbf{e}\right)}{ Pr\left(\textbf{q}\, |\, \textbf{e}\right)}$.

\subsection{Selecting optimal representation}
Section \ref{sec:error_estimation} and \ref{sec:bounds_prob_queries} establishes a method to evaluate error bounds for a given AC in terms of number of bits. Next, \ToolName{} finds the least number of fixed-pt and float-pt bits needed for given requirements. For this, it evaluates the bounds starting with 2 fraction bits and 2 mantissa bits, and increments them until the error-requirement is satisfied. Then, it estimates the least number of integer and exponent bits required by the min and max analysis explained in section \ref{IntExpBits}. In this way, \ToolName{} comes up with the optimal fixed-pt and float-pt representation shown in figure \ref{fig:ToolFlow}. 

Subsequently, the framework has to select among fixed-pt and float-pt. \ToolName{} selects the one with the lowest energy-consumption, estimated using operator-level energy models.
Energy models for the adders and multipliers are developed by synthesizing them with varying fraction/mantissa bits and integer/exponent bits in TSMC 65nm technology and extracting post-synthesis energy consumption. The models were fitted using least-squares method to the simulation results, and are summarized in Table \ref{tab:EnergyModels}. 
\setlength{\abovecaptionskip}{15pt}
\begin{table}[!htbp]
\caption{Energy models for arithmetic operators at 1V. N is the \#fixed-pt bits and M is the \#mantissa bits}
\begin{tabular}{|l|c|}
\hline
\multicolumn{1}{|c|}{\textbf{Operator}} & \textbf{Energy (fJ)} \\ \hline
Fixed-pt add                            & 7.8 N                     \\ \hline
Fixed-pt mult                           & 1.9 N$^2$logN                \\ \hline
Float-pt add                            & 44.74 (M+1)                 \\ \hline
Float-pt mul                            & 2.9 (M+1)$^2$ log (M+1)      \\ \hline
\end{tabular}
\label{tab:EnergyModels}
\end{table}
\setlength{\abovecaptionskip}{5pt}%

\subsection{Automatic hardware generation}
\begin{figure}[!t]
\centering
\includegraphics[width=0.7\columnwidth]{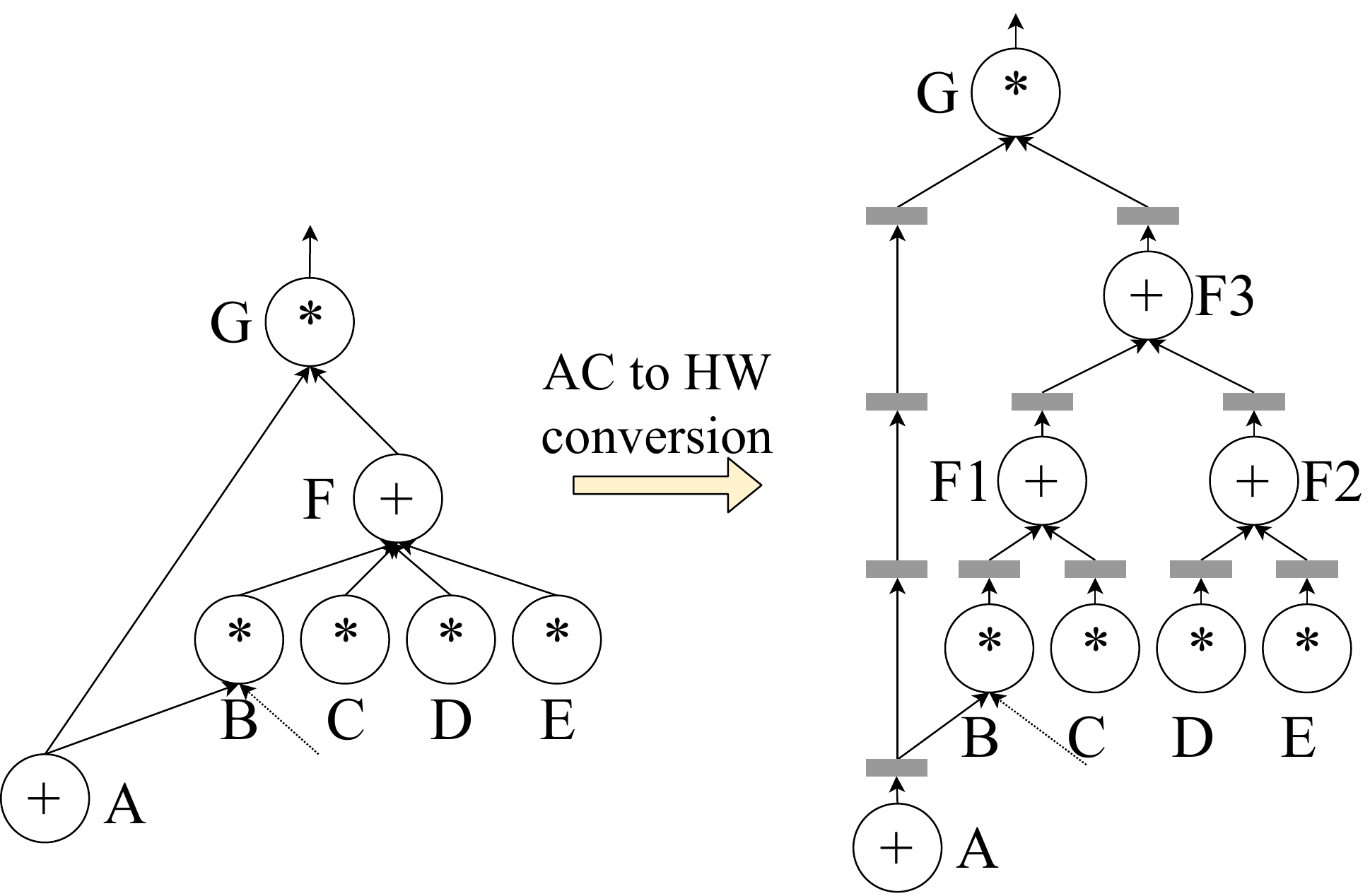}
\caption{Automatic conversion of an AC to a pipelined hardware.} 
\label{fig:HWgeneration}
\end{figure}

\ToolName{} suggests the most-appropriate low-precision representation for the AC, but this may not translate to energy savings unless hardware has custom arithmetic operators. To address this, \ToolName{} has an integrated hardware generator that generates custom parallel hardware that is fully-pipelined and consists of arithmetic operators of the exact precision that is required to meet the user requirements. There are two major stages in the hardware generation process. In the first stage, all AC operators with more than two inputs are decomposed into a tree of 2-input operators. An example of such decomposition is shown in figure \ref{fig:HWgeneration}, wherein the F operator is decomposed into a tree of F1, F2 and F3. In the second stage, the generator inserts pipeline registers after every operator. In some cases, it may have to insert multiple registers due to a mismatch in path timings, as shown in the path between A and G in figure \ref{fig:HWgeneration}. The final output of \ToolName{} is a verilog code of the custom hardware.

\setlength{\abovedisplayskip}{-3pt}
\setlength{\abovedisplayskip}{-3pt}

\begin{table*}[]
\caption{Optimal fixed-pt and float-pt representations that meets error tolerance with selected repr. in bold. Measured error and energy for the selected representation. I,F,E and M stands for number of integer, fraction, exponent and  mantissa bits}
\centering
\begin{tabular}{|c|c|c|c|c|c|c|c|}
\hline
\textbf{AC} & \begin{tabular}[c]{@{}c@{}}\textbf{Type of}\\ \textbf{query}\end{tabular} & \begin{tabular}[c]{@{}c@{}} \textbf{Error} \\ \textbf{tolerance} \end{tabular}& \begin{tabular}[c]{@{}c@{}} \textbf{Opt. Fx-pt repr. I, F} \\ \textbf{ $($pred. energy in} \\ \textbf{nJ/AC\_eval$)$} \end{tabular} & \begin{tabular}[c]{@{}c@{}} \textbf{Opt. Fl-pt repr. E, M} \\ \textbf{ $($pred. energy in} \\ \textbf{nJ/AC\_eval$)$} \end{tabular} & \begin{tabular}[c]{@{}c@{}} \textbf{Max error} \\ \textbf{observed} \\ \textbf{on test-set} \end{tabular} & \begin{tabular}[c]{@{}c@{}}\textbf{Post-synth.}\\ \textbf{energy}\\ \textbf{(nJ/AC\_eval)}\end{tabular} & \begin{tabular}[c]{@{}c@{}} \textbf{Energy of}\\ \textbf{32b Fl-pt} \\ \textbf{E=8, F=23} \end{tabular} \\ \hline

\multirow{4}{*}{HAR}    & Marg. prob. & abs. err 0.01 & \textbf{1, 15 (4.3)} & 9, 14 (6.7) & 5.9x10$^{-4}$   & 5.3 & \multirow{4}{*}{10.8} \\ \cline{2-7} 
                        & Marg. prob. & rel. err 0.01 & 1, >64 ( - ) & \textbf{9, 14 (6.7)} & 1.0x10$^{-3}$   & 7.2 &\\ \cline{2-7} 
                        & Cond. prob. & abs. err 0.01 & 1, >64 ( - ) & \textbf{9, 14 (6.7)} & 2.6x10$^{-4}$   & 7.2 &\\ \cline{2-7} 
                        & Cond. prob. & rel. err 0.01 & -                   & \textbf{9, 14 (6.7)} & 1.0x10$^{-3}$   & 7.2 &\\ \hline
\multirow{2}{*}{UNIMIB} & Marg. prob. & abs. err 0.01 & \textbf{1, 13 (0.4)} & 7, 12 (0.6) & 4.9x10$^{-4}$   & 0.34 & \multirow{2}{*}{0.89} \\ \cline{2-7} 
                        & Cond. prob. & rel. err 0.01 & -                   & \textbf{7, 12 (0.6)} & 1.1x10$^{-3}$   & 0.44 &\\ \hline

\multirow{2}{*}{UIWADS} & Marg. prob. & abs. err 0.01 & \textbf{1, 11 (0.06)}& 6, 10 (0.09)& 1.3x10$^{-3}$ & 0.06 & \multirow{2}{*}{0.18} \\ \cline{2-7} 
                        & Marg. prob. & rel. err 0.01 & 1, 47 (1.3) & \textbf{6, 10 (0.09)}& 1.2x10$^{-3}$ & 0.08 & \\ \hline 
\multirow{2}{*}{Alarm}  & Marg. prob. & abs. err 0.01 & \textbf{1, 14 (2.2)} & 8, 13 (3.2) & 2.2x10$^{-4}$ & 2.43 & \multirow{2}{*}{5.37} \\ \cline{2-7} 
                        & Cond. prob. & rel. err 0.01 & -                   & \textbf{8, 13 (3.2)} & 2.8x10$^{-4}$ & 3.18 &\\ \hline 

\end{tabular}
\label{tab:exp2}
\end{table*}

\section{Experimental results} \label{sec:experiments}
We validate the functionality of \ToolName{} for the Arithmetic circuits targeting embedded sensing applications, by performing two types of experiments on four datasets. Three of these datasets (HAR \cite{anguita2013public}, UNIMIB\cite{app7101101}, UIWADS \cite{casale2012personalization} in table \ref{tab:exp2}) correspond to activity and user identification applications in smartphones and therefore rely on the accurate estimation of a conditional probability of the form $Pr(Activity|sensors)$ to make threshold-based decisions. The fourth dataset (Alarm in table \ref{tab:exp2} \cite{beinlich1989alarm}) is of a patient monitoring application and is often used as a standard Bayesian network benchmark. 

The ACs used in this section are compiled using the ACE tool \cite{darwicheace}, with -cd06 and -forceC2d option enabled. For the experiments on HAR, UNIMIB, and UIWADS, we trained Naive Bayes classifier on 60\% of the data and used the rest for testing. The testing dataset for Alarm is generated by sampling 1000 instances from the trained network. 
In all the experiments, the leaf nodes of the BN were used as evidence nodes \textbf{e} and one of the root nodes in the BN (the class node in the case of the classifiers) as a query node \textbf{q}.

\subsection{Validation of bounds}
This experiment confirms the validity of the derived error bounds for the AC compiled from the Alarm network. The experimental setting is as follows:

\textbf{Fixed-pt}: The number of integer bits is set to 1 based on the max-analysis , and fraction bits is varied from 8 to 40.

\textbf{Float-pt}: The number of exponent bits is set to 8 based on the max-min analysis, and mantissa bits is varied from 8 to 40. 

\begin{figure}[!t]
\begin{subfigure}{0.49\columnwidth}
\centering
\includegraphics[trim={0 0 0 0},clip,width=\columnwidth]{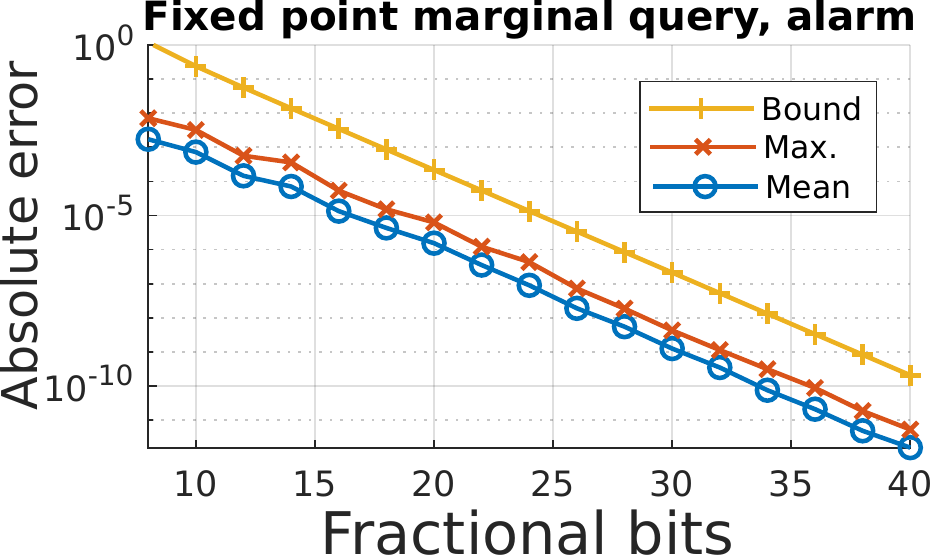}
\caption{ }
\label{}
\end{subfigure}%
\begin{subfigure}{0.49\columnwidth}
\centering
\includegraphics[trim={0 6pt 20pt 8pt},clip,width=\columnwidth]{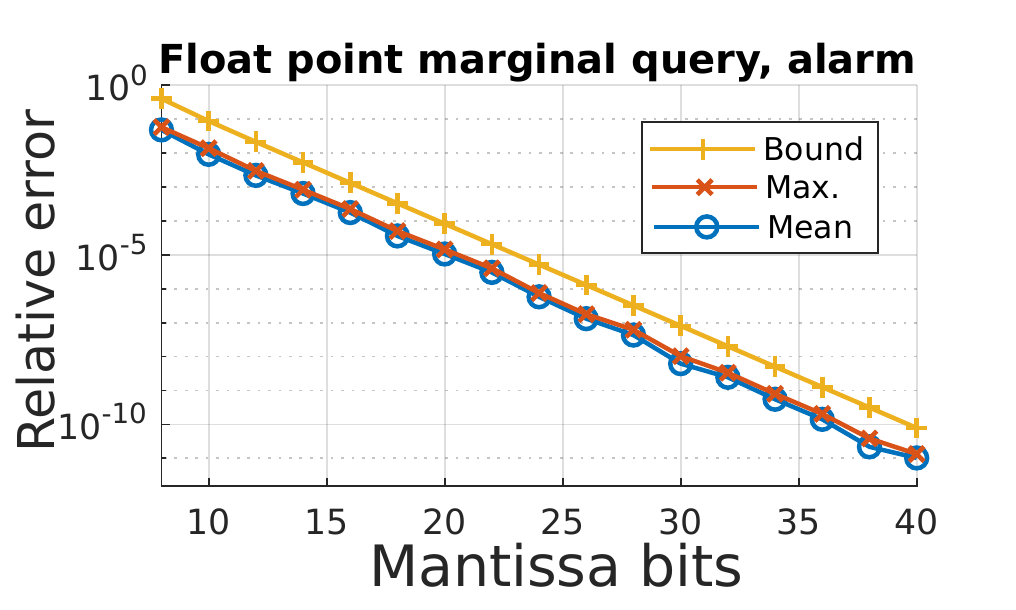}
\caption{ }
\end{subfigure}

\caption{Analytical error bounds and the observed error over a test-set, for varying (a) fraction and (b) mantissa bits, validating the correctness of bounds}
\label{fig:fx_flt_plot}
\end{figure}

Figure \ref{fig:fx_flt_plot} shows the max and mean error on the test-set, which confirm the validity of the bounds.

\subsection{Overall performance}
In this experiment, the complete \ToolName{} framework is deployed to choose an appropriate arithmetic representation and generate hardware for different ACs and for given user requirements. The results of the experiment are summarized in Table \ref{tab:exp2}. Experiments are performed for all combinations of queries and types of error tolerances for the HAR AC, and two combinations for the rest of the ACs. The table shows the optimal fixed-pt and float-pt representation that meet the target error-tolerance. Among these, \ToolName{} selects the one with less predicted energy, highlighted in bold. The resulting maximum error observed on the test-sets remain within the required error-tolerance. The post-synthesis energy consumption matches well to the energy predicted by the framework. The energy consumption of the hardware with a 32b float (E=8, M=23, 1 sign bit) is also shown for comparison. Note here that the choice of 0.01 error tolerance is arbitrary and higher energy-efficiency can be achieved for relaxed error tolerances. 

\setlength{\abovecaptionskip}{10pt}
\setlength{\abovecaptionskip}{5pt}%

\section{Conclusion} \label{sec:conclusion}
Probabilistic inference with Arithmetic circuits can be made energy-efficient by tolerating a small amount of error in output probabilities and by designing custom hardware to exploit this error tolerance. This paper, therefore, proposes \ToolName{}, a holistic framework to automate the design of low-precision custom hardware for ACs. The framework estimates worst-case error bounds for ACs, taking into account the error incurred in reduced precision fixed- and floating-point operators. It estimates the impact of these errors on different types of probabilistic queries and finds the least number of fixed-pt and float-pt bits required to meet the error-tolerance. Subsequently, it chooses among the fixed-pt and float-pt representation based on the energy models developed for this purpose. Next, \ToolName{} automatically converts an AC to pipelined logic with custom arithmetic operators. The analytically derived error bounds are validated for varying fixed- and float-pt bits. Finally, the \ToolName{} framework is used for several embedded sensing benchmarks, confirming that the error-requirements are met and the energy consumption of automatically generated hardware matches the prediction. 

\printbibliography

\end{document}